\documentclass[fleqn,10pt]{wlscirep}

\usepackage{latexsym,epsfig,graphicx}
\usepackage{dcolumn}
\usepackage{bm}
\usepackage{comment}
\usepackage{amsmath,eqnarray,amssymb}

\newcommand{\bma}{\begin{pmatrix}}
\newcommand{\ema}{\end{pmatrix}}
\newcommand{\balig}{\begin{align}}
\newcommand{\ealig}{\end{align}}

\title{Engineering of many-body Majorana states in a topological insulator/s-wave superconductor heterostructure}

\author[2,*]{Hsiang-Hsuan Hung}
\author[1,*]{Jiansheng Wu}
\author[3]{Kuei Sun}
\author[4,5,6,$\dag$]{Ching-Kai Chiu}
\affil[1]{Department of Physics, Southern University of Science
and Technology of China, Shenzhen, Guangdong 518055, P. R. China}
\affil[2]{Department of Physics, University of Texas at Austin,
Austin, Texas 78712-1192, USA} \affil[3]{Department of Physics,
The University of Texas at Dallas, Richardson, Texas 75080-3021,
USA} \affil[4]{Condensed Matter Theory Center and Joint Quantum
Institute and Maryland Q Station, Department of Physics,
University of Maryland, College Park, MD 20742-4111, USA}
\affil[5]{Department of Physics and Astronomy, University of
British Columbia, Vancouver, BC, Canada V6T 1Z1} \affil[6]{Quantum
Matter Institute, University of British Columbia, Vancouver BC,
Canada V6T 1Z4}

\affil[*]{Authors with equal contribution}
\affil[$\dag$]{Corresponding author. Email:
 \href{mailto:chiu7@umd.edu}{chiu7@umd.edu}}

\begin{abstract}
\textbf{We study a vortex chain in a thin film of a topological
insulator with proximity-induced superconductivity---a promising
platform to realize Majorana zero modes (MZMs)---by modeling it as
a two-leg Majorana ladder. While each pair of MZMs hybridizes
through vortex tunneling, we hereby show that MZMs can be
stabilized on the ends of the ladder with the presence of tilted
external magnetic field and four-Majorana interaction.
Furthermore, a fruitful phase diagram is obtained by controlling
the direction of magnetic field and the thickness of the sample.
We reveal many-body Majorana states and interaction-induced
topological phase transitions and also identify
trivial-superconducting and commensurate/incommensurate
charge-density-wave states in the phase diagram.}
\end{abstract}
\begin{document}

\flushbottom

\maketitle

\thispagestyle{empty}

\section*{Introduction}

The exploration of various symmetry-protected topological states
in quantum systems has become an intensively focused field in
condensed-matter and AMO physics~\cite{Kane2010,
Zhang2011,chiu_review15}. Quantum matter hosting Majorana zero
mode (MZM), a particle being its own antiparticle, is of
particular interest in the research forefront for its capability
of revealing the intriguing nature of quantum entanglement and
performing fault-tolerant quantum computation~\cite{Wilczek2009,
Fu2008, Das2010, Mourik2012, Alicea2013, Sato2009, Kitaev2001,
Kitaev2003, Nayak2008}. Recently, a pair of Majorana fermions in a
one-dimensional (1D) system has been theoretically proposed and
experimentally implemented in a semiconductor nanowire or a
magnetic-atom chain on a superconducting substrate, producing an
ideal quantum qubit~\cite{Kitaev2001, Das2010, Sau2010}. However,
efficient quantum information processing requires multiple qubits
that can be practically manipulated. For this purpose, a more
attractive candidate is the heterostructure of a three0dimensional
(3D) topological insulator (TI) film and an s-wave superconductor,
which can carry a vortex array with a pair of MZMs embedding in
each vortex and localizing around the top and bottom surfaces of
the film, respectively~\cite{Fu2008}. The proximity effect of
superconductivity has been confirmed that the superconductivity on
the naked surface of the TI film is induced from the other side of
the TI
surface~\cite{Xu_TI_SC,wang_Xue_science_12,Majorana_Fu_Kane_Jia,SC_Proximity_Jia}
in contact with the superconductor as illustrated in
Fig.~\ref{fig:leg_model} (a). In experiments, MZMs can be observed
only on the naked surface since the interface between TI and the
superconductor has been buried. Currently, the observation of
zero-bias conductance peak and spin selective Andreev reflection
in the vortices shows the tentative evidence of
MZMs~\cite{Majorana_Fu_Kane_Jia,PhysRevLett.116.257003}.

Although the zero bias peak has been observed in the vortex cores
of the naked TI surface, existence of the MZMs remains debatable
in current heterostructure experiments due to two major issues.
First, MZM and the low-energy Caroli-de-Gennes-Matricon mode
\cite{Caroli_mode} ($\sim 0.01$meV) embedding in the vortex are
indistinguishable due to the current energy resolution ($\sim
0.1$meV) in scanning tunneling spectroscope.  Second, the TI
should be thin enough such that the superconductivity can be
proximity-induced on the naked TI
surface~\cite{Chiu_SC_induced_gap} but should be thick enough to
suppress the Majorana hybridization on the top and bottom TI
surfaces. In the recent experiment~\cite{Majorana_Fu_Kane_Jia},
the thickness ($\sim 5$nm) of TI  causes the order of $1$meV of
the Majorana hybridization. By comparing with the superconducting
gap ($1$meV), this hybridization completely destroys MZMs. To save
MZMs in this experimental setup, first we consider a 1D dense
vortex array in the thin TI film and tune the chemical potential
right at the Dirac point of the surface modes so that additional
chiral symmetry is preserved.  The symmetry suppresses the
hybridization of MZMs on the same surface to zero. Hence, the
interaction of four Majoranas becomes leading
order~\cite{Chiu_Majorana_interaction,Chiu_Majorana_interaction_platform,Chiu_Majorana_interaction_detail},
so many-body Majorana wavefunctions have to be considered for the
full characterization of the system's quantum phases. Furthermore,
when the vortex array is tilted by a magnetic field, the Majorana
interaction assists a MZM to appear on the end of the vortex
array. Such a many-body effect, though it was less investigated
previously, not only provides additional degrees of freedom to
engineer MZMs but also open an avenue to study interacting
topological physics.

In this report, we propose a possible realization of a
one-dimensional vortex array in a superconducting TI film device
that can be represented by a tilted ladder model of many Majorana
fermions associated with the Fu-Kane model~\cite{Fu2008}, as shown
in Fig.~\ref{fig:leg_model} (b). In this system, various
intravotex and intervortex couplings between Majorana fermions are
tunable with the control of the chemical potential and the
vortex's incline angle by an external magnetic field. Performing
the density-matrix-renormalizaion-group (DMRG)
calculations~\cite{White1992, White1993,
Schollwock2005,Shibata2003}, we obtain the many-body ground state
of the system and present interacting phase diagrams as a function
of these Majorana couplings. The presence of Majorana interaction
enlarges the topological region of the Majorana ladder in the
phase diagram; it leads to a MZM localized on the end of the
ladder even in the presence of the Majorana hybridization.

\section*{Results}

\begin{figure*}[t]
\centering
\epsfig{file=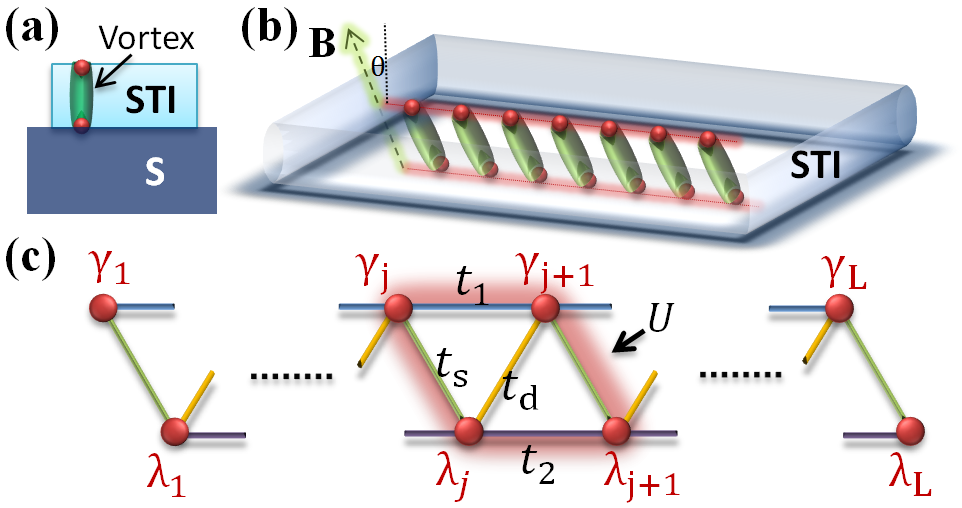,clip=0.1,width=.8\linewidth,angle=0}
\caption{{ Experimental setup for a Majorana ladder.} (a)
Illustration of a pair of Majorana fermions (red dots) embedding
respectively in the top and bottom of a vortex line (green tube)
in a strong-topological-insulator thin film (STI, light-blue
region) on an $s$-wave superconductor (S, dark-blue regions). (b)
An array of vortices in the heterostructure, forming a two-leg
Majorana ladder. An applied magnetic field ${\bf B}$ (dashed
arrow) lines up vortices with angle $\theta$ and hence tilts the
ladder system. (c) The tight-biding model for the Majorana ladder,
describing top-chain ($\gamma$) and bottom-chain ($\lambda$)
Majorana fermions coupled by intra-leg tunnelings $t_1 \gamma_i
\gamma_{i+1}$ and $t_2 \lambda_i \lambda_{i+1}$, inter-leg
tunnelings $t_s \gamma_j \lambda_j$ and $t_d \lambda_j
\gamma_{j+1}$, and four-Majorana interaction $U\gamma_j \lambda_j
\lambda_{j+1} \gamma_{j+1}$. } \label{fig:leg_model}
\end{figure*}

{\large \bf Experimental setup of a two-leg Majorana ladder.} We
start from the Fu-Kane heterostructure~\cite{Fu2008}, which is a
3D strong TI thin film on the top of an s-wave type-II
superconductor. In this thin film, both top and bottom TI surfaces
exhibit effective time-reversal-symmetric $p\pm ip$
superconductivity, via the superconducting proximity effect.
Experimentally this setup has been demonstrated in Bi$_2$Te$_3$
thin films grown on a NbSe$_2$
substrate~\cite{Majorana_Fu_Kane_Jia,SC_Proximity_Jia}, as shown
in Fig.~\ref{fig:leg_model}(a).  With an external magnetic field
turned on, vortices are generated on the TI surfaces and each end
of the vortices hosts a MZM~\cite{Chiu2011, Hung2013,Ashvin_SCTI}.
However, the induced superconducting gap on the naked (top)
surface is much smaller than the bottom surface in contact with
the
superconductor~\cite{SC_Proximity_Jia,Xu_TI_SC,Chiu_SC_induced_gap}.
Furthermore, the MZMs at the two ends of the vortex can tunnel
through the vortex line and then hybridize, such that they do not
possess zero energy. For this purpose, we consider a tilted
magnetic field, which can effectively enlarge the distance between
the MZM on the top and bottom surfaces, and weaken the
hybridization.

Inspired by the one-dimensional vortex chain with a tilted
magnetic field in the copper oxide thin films~\cite{Henini2001},
we consider a strongly anisotropic vortex array, which turns out
to be a one-dimensional stripe along a certain direction
determined by an external magnetic field, as shown in
Fig.~\ref{fig:leg_model}(b). With the tilted fields, the MZMs (red
dots) at the top and bottom surfaces oppositely shift and form a
tilted two-leg ladder, as shown in Fig.~\ref{fig:leg_model}(c). On
the same surface, the wavefunction of the MZM may overlap with its
nearest neighbors and contribute to intra-leg hopping $t_1\gamma_j
\gamma_{j+1}$ and $t_2 \lambda_j \lambda_{j+1}$ for the top and
bottom surfaces, respectively. The thickness of the TI determines
the coupling between the top and bottom Majoranas along the same
vortex line, $t_s \gamma_j \lambda_j$. As the magnetic field is
titled, the hybridization between $\gamma_{j+1}$ and $\lambda_{j}$
becomes non-negligible, resulting in $t_d \gamma_{j+1}\lambda_j$.
In addition to the single-particle hopping, there exists
interaction among the MZMs. Assuming that the tilted angle
$\theta$ is small enough such that the distance between $\gamma_j$
and $\lambda_j$ is less than that between $\gamma_{j+1}$ and
$\lambda_j$, at the leading order we can have interaction stemming
from four neighboring Majoranas in a closed loop
$\gamma_j\lambda_j \lambda_{j+1}\gamma_{j+1}$. Thus, the whole
Hamiltonian in the Majorana representation reads
\begin{align}
\hat{H}_{\rm{M}}= i\sum^{L-1}_{j=1} \big ( t_1
\gamma_j\gamma_{j+1} + t_2 \lambda_j \lambda_{j+1}  +t_d
\gamma_{j+1}\lambda_j \big) + i t_s \sum^L_{j=1} \gamma_j
\lambda_j + U \sum_{j=1}^{L-1} \gamma_j \lambda_j \lambda_{j+1}
\gamma_{j+1}, \label{Majorana basis}
\end{align}
where $L-1$ in the first summation indicates the open boundary
condition. One feasible way to control $t_{1,2}$ is to adjust the
spatial distance between vortices, which can be artificially tuned
via the magnitudes of magnetic fields; at the same time, however,
the four-Majorana interaction is weaken. To keep the interaction
strength, one needs to tune the chemical potential at the surface
Dirac point to preserve additional chiral symmetry. The Majorana
hybridization on the surface, which is forbidden by the symmetry,
vanishes, and the Majorana interaction, which preserves the
symmetry,
survives~\cite{Chiu_Majorana_interaction,Chiu_Majorana_interaction_platform,Chiu_Majorana_interaction_detail}.

On the other hand, in the noninteracting limit, $U=0$,  the
topology of the Majorana ladder is determined by $t_d/t_s$. The
topological region where MZM reside on the vortex cores can be
exactly determined at $|t_d/t_s| > 1$ (See {\bf Method:
Noninteracting Majorana ladder}). To solve the finite-$U$ cases,
one can transform the Majorana Hamiltonian Eq.~(\ref{Majorana
basis}) in terms of conventional fermionic operators
$\gamma_j=c_j+c^{\dag}_j$ and $\lambda_j = i (c_j - c_j^\dagger)$
as
\begin{align}
\hat{H}_{\rm{F}} =\sum^{L-1}_{j=1} \Big( \bar{t}c^{\dag}_j
c_{j+1}^{} + \Delta c^{\dag}_j c^{\dag}_{j+1} +h.c. \Big) - (2t_s
+4U)\sum^L_{j=1}n_j +2U(n_1+n_L) +4U\sum^{L-1}_{j=1}n_j n_{j+1}
,\label{fermionic basis}
\end{align}
where $n_j=c^\dagger_jc_j^{}$, $\bar{t}=-t_d+i(t_1+t_2)$, and
$\Delta = -t_d+i(t_1-t_2)$. Here we drop the constant energy shift
after the basis transformation. We also consider a grand canonical
ensemble such that the filling of fermions changes with the
on-site term $-(2t_s+4U)\sum_j n_j$.

The spinless fermionic Hamiltonian (\ref{fermionic basis}) has a
similar structure to an interacting Kitaev
chain~\cite{Kitaev2001,Thomale2013,Chan2015}. However, we should
emphasize that the realistic system (a Majorana ladder in a vortex
chain) described by our model is fundamentally different from that
in the previous study. Our proposed heterostructure provides a
very different mechanism of tuning the model parameters, enabling
the exploration of a wider phase diagram. For example, the last
term, which has the form of nearest-neighbor electronic
interaction, is actually determined by the overlap between four
Majorana fermions in a plaquette $\gamma_j\lambda_j
\lambda_{j+1}\gamma_{j+1}$. Therefore the strength $U$ is related
to the sample thickness and distance between two vortices on the
same surfaces and can hence be fine tuned (compared with the
hardly tunable electronic interaction in solid). To capture the
salient physics, below we consider non-negative tight-binding
parameters and interaction strength, i.e.~$t_{1,2,s,d} \ge 0$ and
$U \ge 0$, the same intra-leg tunnelings on the top and bottom
surfaces $t_1=t_2$, and the inter-leg hopping $t_s=1$ as the
energy unit. Moreover, we are interested in the phases of the
entire ladder, which should not be sensitive to the boundary
condition, so we neglect the boundary term $2U(n_1+n_L)$ in
Eq.~(\ref{fermionic basis}) in following calculations.

Before getting into the details of the system's phase diagram, we
briefly point out that the original Hamiltonian of
Eq.~(\ref{Majorana basis}) has another equivalent form in terms of
Pauli matrices $\sigma_{x,y,z}$ as $\hat{H}_{\rm{S}}=\sum_j \big (
-t_s \sigma_j^x - t_1 \sigma_j^y \sigma_{j+1}^z + t_2 \sigma_j^z
\sigma_{j+1}^y + t_d \sigma_j^z \sigma_{j+1}^z + U \sigma_j^x
\sigma_{j+1}^x \big )$, which can be obtained through a
Jordan-Wigner transformation $\gamma_j = (\prod_{k}^{j-1}
\sigma_k^x) \sigma_j^z$ and $\lambda_j = -(\prod_{k}^{j-1}
\sigma_k^x) \sigma_j^y$. This Hamiltonian describes a spin chain
with transverse Zeeman field ($t_s$), anisotropic
Dzyaloshinskii-Moriya interaction
($t_{1,2}$)~\cite{Interactation_Ising,PhysRevLett.4.228}, and
anisotropic exchange interaction ($t_d$ and $U$). Our proposed
heterostructure may thus find applications as a test bed to such
interesting spin systems.

\begin{figure*}[t]
\centering
\epsfig{file=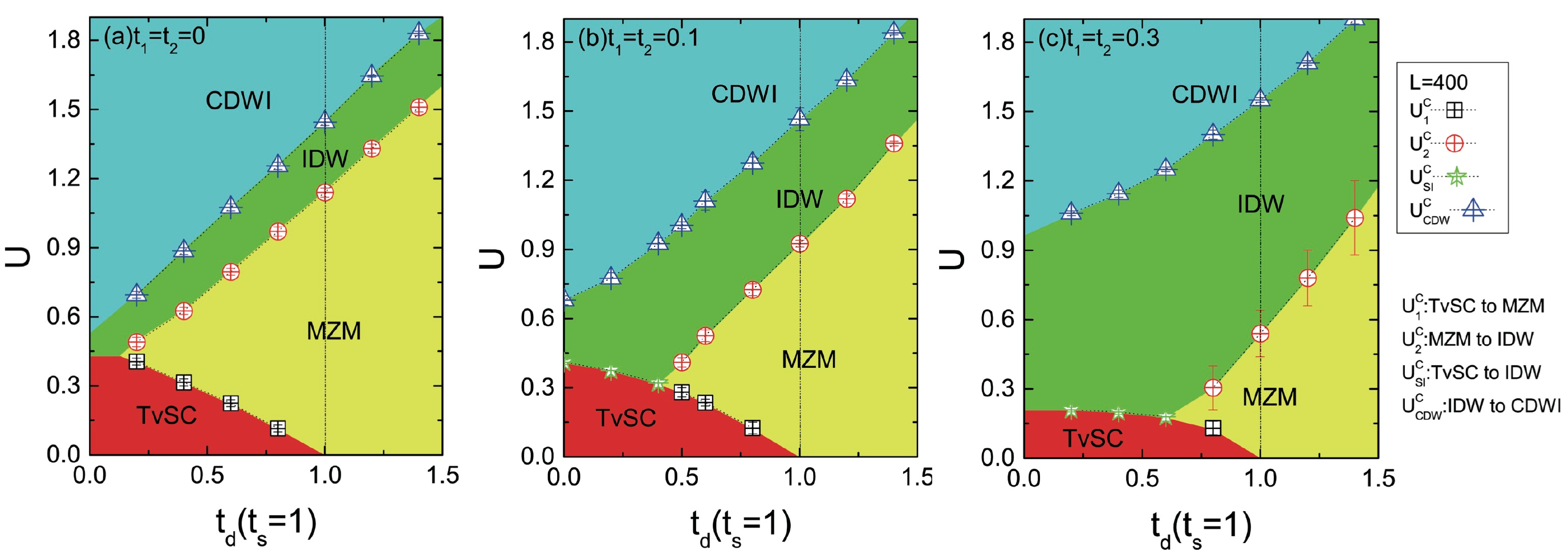,clip=0.1,width=1\linewidth,angle=0}
\caption{{Phase diagram for various $t_{1,2}$ values. } The $t_d$
vs $U$ phase diagram of the interacting Majorana ladder at (a)
$t_{1,2}=0$ (b) $t_{1,2}=0.1 t_s$ and (c) $t_{1,2}=0.3 t_s$. The
dot line locates the noninteracting topological phase boundary
$t_d=t_s$. TvSC, MZM, IDW and CDWI separately represent the
trivial superconducting states, Majorana zero modes,
incommensurate charge-density-wave liquids and commensurate
charge-density-wave insulators. Phase boundaries between TvSC and
MZM, MZM and IDW, TvSC and IDW, and IDW and CDWI, are described by
black squares (corresponding to critical interaction $U^c_1$), red
circles ($U^c_2$), green stars ($U^c_{ \rm {SL}}$), and blue
triangles ($U^c_{ \rm{CDW}}$), respectively.}
\label{fig:phasediagram}
\end{figure*}

{\large \bf Phase diagram}. The Hamiltonian Eq.~(\ref{fermionic
basis}) is effectively a 1D fermion chain. To study the many-body
physics, we implement the DMRG method to perform numerical
simulation, and investigate the ground state phase diagram. We
compute the energy gap defined as the difference of the ground
state energy in the even parity ($P=1$) and odd parity ($P=-1$)
sectors $\Delta E=|E_0(P=+1)-E_0(P=-1)|$, the difference in paired
entanglement spectra $\delta \varepsilon$ ($\delta \varepsilon=0$
indicates two-fold degeneracy of entanglement spectrum), charge
structure factor $S(q)$ (which indicate the strength of charge
density wave with momentum $q$) and filling $\bar{n}$ of fermions
(see {\bf Method: Physical quantities} for the details). There are
several distinct phases such as trivial superconducting phase
(TvSC), topological Majorana zero mode (MZM), incommensurate
charge-density-wave liquid (IDW) and commensurate
charge-density-wave insulators (CDWI) depending on the system
parameters.  We summarize the phase diagram as a function of $t_d$
and $U$ and at a variety of $t_{1,2}$ in
Fig.~\ref{fig:phasediagram}.


\begin{figure*}[t]
\centering
\epsfig{file=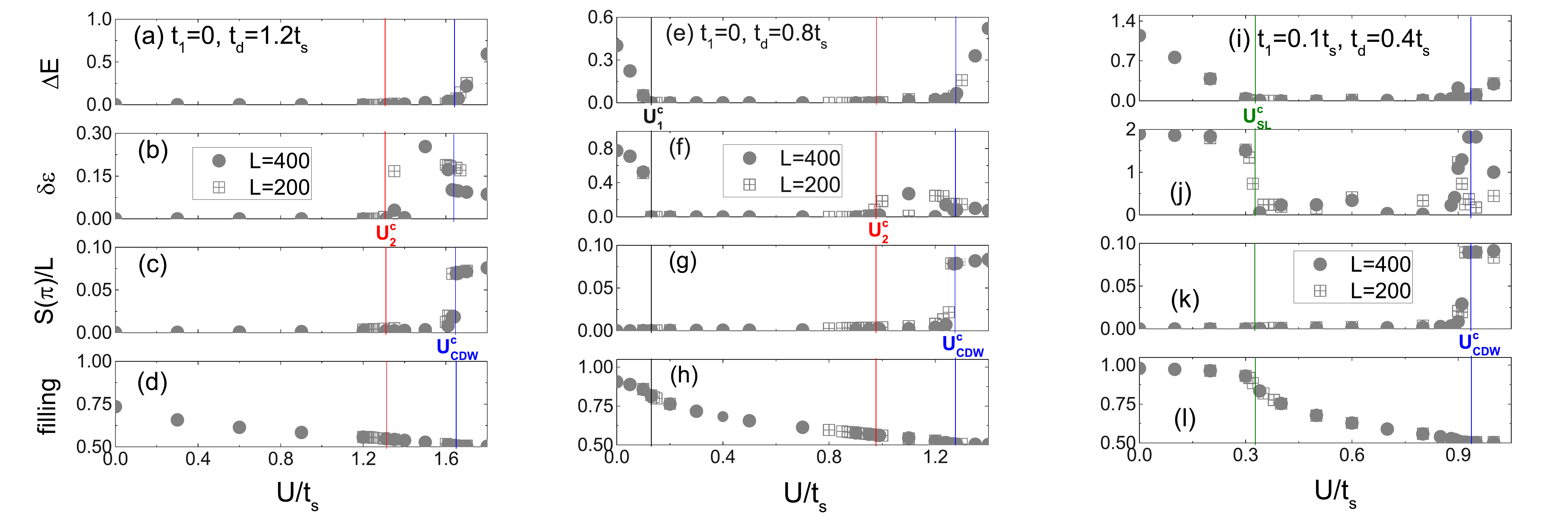,clip=0.1,width=1\linewidth,angle=0}
\caption{{Key physical quantities vs $U$. } From top to bottom
rows: energy gap $\Delta E$, quantity characterizing
entanglement-spectrum degeneracy $\delta \varepsilon$, charge
structure factor at $q=\pi$ (per length) $S(\pi)/L$, and filling
of fermions $\bar{n}$, respectively. From left to right columns:
(a-d) $t_{1,2}=0$ and $t_d=1.2 t_s$, (e-h) $t_{1,2}=0$ and
$t_d=0.8 t_s$, and (i-l) $t_{1,2}=0.1 t_s$ and $t_d=0.4 t_s$,
respectively.} \label{fig:measurements}
\end{figure*}

First let us simply consider the $t_{1,2}=0$ case as the chemical
potential is adjusted at the surface Dirac node, i.e.~the
intra-leg hopping vanish, in Fig.~\ref{fig:phasediagram}
(a)~\cite{Chiu_Majorana_interaction}.  In this case, the vortices
on the same TI surfaces far separate in space. The noninteracting
MZM exists as $t_d>t_s$. As the interaction is slightly turned on,
we found that the ladder is still in the topological phase. Since
the bulk-boundary correspondence still holds in interacting class
D, a many-body MZM is localized on each end of the ladder for
non-zero $U$ \cite{Chiu_Majorana_interaction_detail}. This
many-body MZM is adiabatically connected to the single-particle
MZM without the interaction. In Fig.~\ref{fig:measurements}(a-d),
we show details of four physical quantities vs $U$ at fixed
$t_d=1.2 t_s$ to identify the topological phase. The ground state
energy in the even parity ($P=+1$) and odd parity ($P=-1$) sectors
are doubly degenerate, so $\Delta E=0$ in
Fig.~\ref{fig:measurements}(a). Furthermore, in Fig.
\ref{fig:measurements}(b) $\delta \varepsilon=0$, which indicates
double degeneracy in the entanglement spectra, leads to the
topological phase extended from $U=0$. Upon increasing $U$, on the
other hand, the smoothly decreasing filling $\bar{n}$ and the
absence of featured peaks in $S(q)$ show no indication of other
physical phases. {\color{red}}

\begin{figure*}[t]
\centering
\epsfig{file=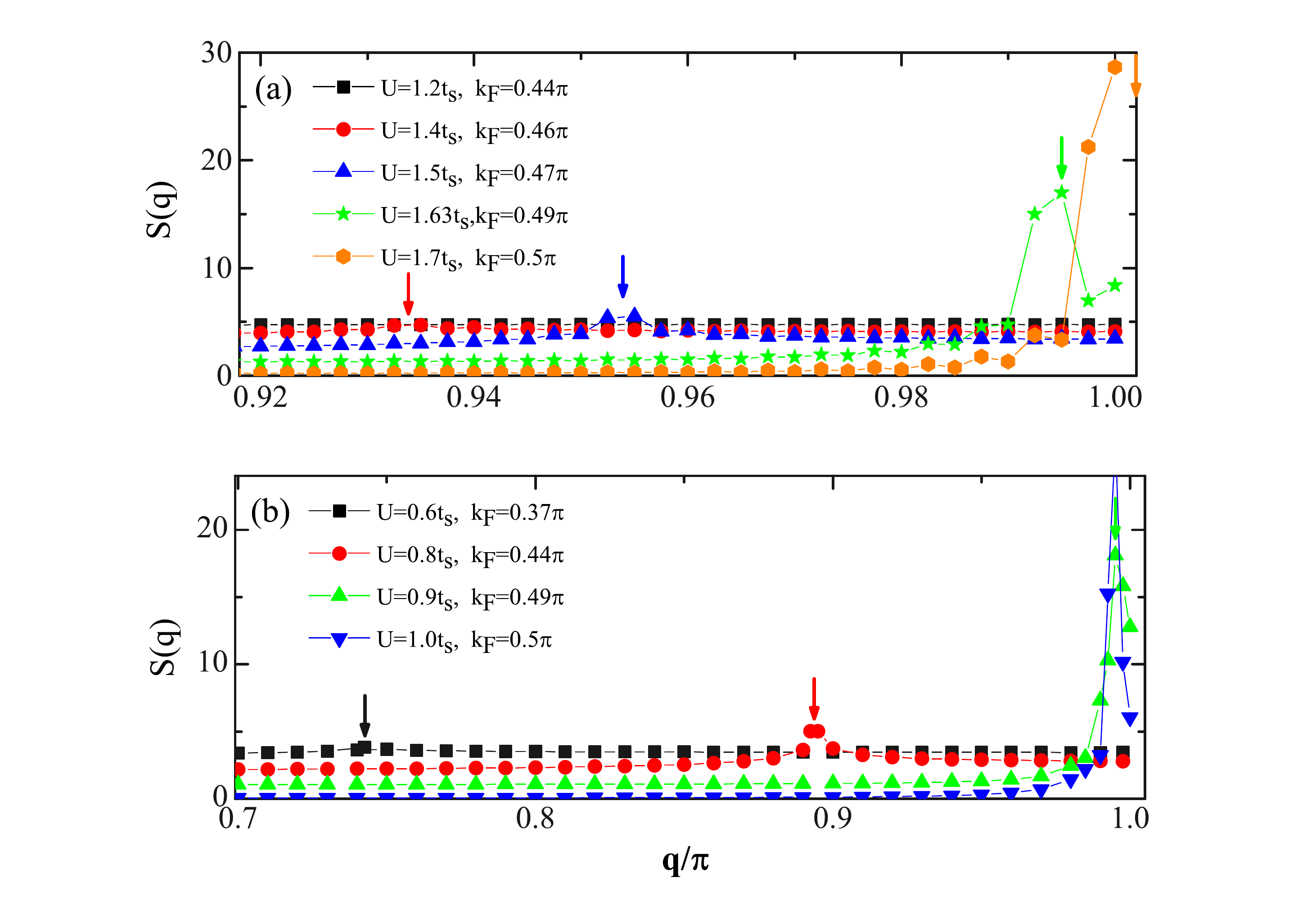,clip=0.12,width=0.8\linewidth,angle=0}
\caption{ {Charge structure factor $S(q)$ of the Majorana ladder.
} The structure factor of the charge-charge correlation on the
interacting Majorana ladder at (a) $t_{1,2}=0$, $t_d=1.2t_s$ and
(b) $t_{1,2}=0.1 t_s$, $t_d=0.4$ and various interaction strength
(data are vertically shifted in arbitrary units for a clear view).
The size is $L=400$. The arrows point out the location of peaks.}
\label{fig:chargestructure}
\end{figure*}

The MZM phase region is extended as interaction $U$ increases
until the phase transition $U^c_2$ [red circles in Fig.
\ref{fig:phasediagram} (a) ] to the incommensurate
charge-density-wave liquid (IDW). To show the IDW region, we can
see the double degeneracy between the even- and odd-parity ground
states is clearly lifted~\cite{Thomale2013} (even if the energy
gap is quite small) in Fig.~\ref{fig:measurements}(a). Meanwhile,
as shown in Fig.~\ref{fig:measurements}(b) the double degeneracy
of entanglement spectrum disappears, i.e.~$\delta \varepsilon
>0$. The charge structure factor $S(q)$ shows peaks at the
incommensurate wave vector at $q\cong 2k_F$, where $k_F$ is the
Fermi vector. An example can be seen
Fig.~\ref{fig:chargestructure} (a) that the charge structure
factors $S(q)$ at different interaction strength as $t_{1,2}=0$
and $t_d=1.2t_s$. In this regime, filling $\bar{n}$ still
decreases smoothly upon increasing $U$, and the Fermi vector $k_F$
as well as the peak locations of $S(q)$ move towards to a larger
$q$. This charge $2k_F$ instability of IDW state is also
reminiscent of a similar feature of a Luttinger
liquid~\cite{giam04,RevModPhys.84.1253}. In
Fig.~\ref{fig:phasediagram}(a), the red circles describe the phase
boundary between MZM and IDW.

As $U$ increases across the other phase boundary [blue triangles
in Fig. \ref{fig:phasediagram}(a) or blue line in Fig.~
\ref{fig:measurements}(c)], the system opens a gap and a CDWI is
detected. The dominant peak occurs at $q=\pi$ and the CDW order
parameter survives in the thermodynamic limit. At this moment, the
filling approaches $\bar{n}\simeq 0.5$ or half-filling. The ground
state is parity odd ($P=-1$) and $\Delta E\ne 0$. In a classical
analogy, electrons are loaded on every other lattice site. The
blue triangles ($U_{\rm{CDW}}$) depict the phase boundary between
IDW and CDWI. By DMRG, we can distinguish the distinct phases and
pin out the phase boundary by observing variations in $\Delta E$,
$\delta \varepsilon$ and  $S(\pi)$. The phase transition between
IDW and CDW was also discovered theoretically by bond entropy
method~\cite{Schmitteckert2008}. Experimentally, the appearance of
CDWI can be measured by Coulomb drag~\cite{Laroche2014} or by
thermodynamics method~\cite{Jonson1997}.

Next we turn to the $t_d<t_s$ regime, which physically corresponds
to small tilted angle $\theta$ in Fig.~\ref{fig:leg_model}(b). We
used $t_d=0.8t_s$ as demonstration presented in
Fig.~\ref{fig:measurements}(e-h).  In the noninteracting limit
($U=0$), the system is a trivial superconductor (TvSC) with a
finite gap, because the Majorana hybridization $t_s$ between the
top and bottom TI surfaces destroys the topological phase. The
ground state is parity even ($P=1$) and the entanglement spectrum
shows no paired degeneracy, so both $\Delta E, \delta \varepsilon
\ne 0$ at small $U$ in Fig.~\ref{fig:measurements}(e,f). However,
at a sufficiently strong (but not too strong) interaction
strength, the ladder undergoes the topological phase transition at
$U^c_1$, and MZMs emerge at each end of the ladder. The ground
state has double degeneracy and the entanglement spectrum appears
in pair, i.e.~$\Delta E=\delta\varepsilon=0$. Back to the phase
diagram Fig. \ref{fig:phasediagram} (a), we can clearly see that
the topological state is adiabatically connected to the MZM in the
$t_d >t_s$ regime. This MZM is driven by finite interactions, as
an interaction-induced topological state. The black squares in
Fig.~\ref{fig:phasediagram} (a) describe the phase boundary
between TvSC to MZM (our DMRG calculation shows a weak finite-size
effect on the phase boundary). Upon increasing interaction
strength, the MZM region is enlarged, implying that a moderate
interaction stabilizes the topological MZM, even with less tilted
magnetic fields.  In the large-$U$ side, the ground states are
still characterized as the IDW and CDWI, similar to the
observation in the $t_d>t_s$ regime.

Next we move to consider $t_{1,2} \ne 0$ as the chemical potential
is not located at the surface Dirac node. In reality, TI
materials, that have chemical potential exactly at the Dirac node,
have not been discovered, so the intra-leg tunneling between the
MZMs is inevitable. Therefore it is important to investigate how
MZM responds to finite $t_{1,2}$. The phase diagrams in
Fig.~\ref{fig:phasediagram} (b) and (c) consider finite values of
$t_{1,2}$.  The influence of $t_{1,2}$ is remarkable in the
interacting Majorana ladder.  In Fig. \ref{fig:phasediagram} (b)
using $t_{1,2}=0.1 t_s$, it is obvious to see that, compared to
(a), where $t_{1,2}=0$, the MZM regime shrinks. However, this
phase still extends to a finite range whereas the IDW regime is
enlarged. There exists a critical $t_d$ to harbor the
interaction-driven MZM, which is $t^c_d \sim 0.5 t_s$. Below this
point, the TvSC phase directly turns to the IDW state, and the MZM
disappears. Both TvSC and IDW show trivial behavior in the
entanglement spectra. To pin out the boundary, indicated by green
stars in Fig.~\ref{fig:phasediagram}(b), we examine the gap
magnitudes and observe $S(q)$. At TvSC, $\Delta E\ne 0$ and no
featured peak in $S(q)$, whereas at IDW, $\Delta E \simeq 0$ and
$S(q)$ has a peak located at $q=2k_F$, as shown in Fig.
\ref{fig:chargestructure} (b).  We summarize the variation of the
physics observables at $t_d=0.4 t_s$ and $t_{1,2}=0.1 t_s$ and at
variety of $U$ in Fig. \ref{fig:measurements}(i-l). At $U=0$,
there is an energy gap. The gap $\Delta E$ decreases to a small
but finite value as $U$ increases to $U^c_{SL}$ and remains small
in $U^c_{SL} \le U \le U^c_{CDW}$. At $U>U^c_{CDW}$, $\Delta E$
rapidly increases and $S(\pi)$ jumps to a finite value. In the
whole range $\delta \varepsilon \ne 0$, so no MZM exists.

For further stronger intra-leg tunneling, the region of many-body
MZMs becomes even smaller. Figure \ref{fig:phasediagram} (c) shows
$t_{1,2}=0.3 t_s$. The critical $t_d$ is estimated at $t^c_d \sim
0.8t_s$. Therefore, the presence of intra-leg tunneling $t_{1,2}$
will corrupt the stability of the many-body MZM. We have
numerically estimated that, as $t_1 =t_2 \gtrsim 0.5t_s$, the
system no longer supports the interaction-driven MZM. This implies
that the chemical potential has to be tuned to close to the
surface Dirac node to suppress the intra-leg tunnelings. In
experiment, the magnitudes of magnetic fields are required to be
appropriately tuned, such that the vortices are away from each
other to lower the intra-leg hybridization but close enough to
strengthen the four-Majorana interaction.

\section*{Discussion}

With proper strength of the Majorana interaction, the topological
region is tremendously enlarged; by tilting a small angle of
magnetic field MZMs appear on the ladder ends, even in the
presence of the Majorana hybridization between the top and bottom
surfaces. As shown in Fig.~\ref{fig:phasediagram} the intra-leg
tunneling $t_{1,2}$ of Majorana Fermions on the same surface
shrinks the topological region. To enlarge the region, $t_{1,2}$
can be tuned to zero by adjusting the chemical potential right at
the surface Dirac node. For the recent experiment of the
heterostructure on Bi$_2$Se$_3$ thin films~\cite{Xu_TI_SC}, we
estimate the hybridization strength $t_{d,s}\sim 2$~meV and the
interaction strength $U\sim 0.56$~meV so the ratio $U/t_s \sim
0.28$ (see \textbf{Method: Estimation of Majorana coupling and
interaction}). Hence, as shown in Fig.~\ref{fig:phasediagram} (a),
we can simply tilt the magnetic field such that $t_d/t_s > 0.6$ to
expect the MZM on the end of vortex array of the naked surface.
Our current proposal directly solves one of the major difficulties
of the Fu-Kane model: usually the TI film has to be thin enough to
induce the superconductivity gap on the naked surface, but such a
thin film can lead to the Majorana hybridization, which destroys
MZMs. Tilting the magnetic fields can both reduce the
hybridization and enhance the interaction and hence rescue MZMs.

To explore other many-body phases, such as IDW and CDWI, one needs
other TI materials to provide larger values of $U/t_s$. It is
interesting to see the transition between IDW and MZM, which only
occurs with Majorana interactions. Such a topological phase
transition is beyond the single-particle picture. The IDW state
sharing similarities with a Luttinger liquid could be identified
using the Coulomb drag measurement~\cite{Laroche2014}.

Although the physics of many-body MZM and its topology has been
discussed
extensively~\cite{Fidkowski:2010kx,One_interaction_Kuei,Kells2015,OBrien2015,Katsura2015,Gergs2016},
promising platforms for such systems are barely found in the
literature. In this report, we have designed a realizable
experimental setup to investigate interaction effects on
topological states.

\section*{Method}
{\large \bf Noninteracting Majorana ladder.} To determine the
topological phase, we solve the ladder Hamiltonian
$\hat{H}_{\rm{M}}$ (\ref{Majorana basis}) as the interaction is
off ($U=0$) in the periodic boundary condition by extending the
first summation to $L$ and letting site $L+1$ coincide with site
$1$. By performing Fourier transformation $\gamma_j=\frac{1}{
\sqrt{L}}\sum_k \gamma_ke^{ijk},\ \lambda_j=\frac{1}{
\sqrt{L}}\sum_k \lambda_ke^{ijk}$, the noninteracting Majorana
Hamiltonian in momentum basis is given by
\begin{align}
\hat{H}_{\rm{non}}=\sum_{k} \frac{i}{2} \bma \gamma^\dagger_k &
\lambda^\dagger_k \ema \bma
2i t_1 \sin k & t_s + t_d e^{ik} \\
-t_s -t_d e^{-ik} & 2i t_2 \sin k \\
\ema \bma
\gamma_k \\
\lambda_k \\
\ema. \label{noninteracting Kitaev chain}
\end{align}
The topology of the Majorana ladder can be characterized by the
Pfaffian of the Hamiltonian at $k=0$ and $\pi$
\begin{align}
 (-1)^\nu =
\rm{sgn} \big( \rm{pf} B(0) \rm{pf} B(\pi)\big)= \rm{sgn}\big
((t_s+t_d)(t_s-t_d)\big), \nonumber
\end{align}
where $B(k)=t_s +
t_d e^{ik}$. Hence, the topological region is located at $|t_d|>
|t_s|$ in this non-interacting system, irrespective of values of
$t_1$ and $t_2$. We also expect $t_1$ and $t_2$ small enough to
keep the system insulating. In the topological phase, the MZMs
reside in the vortex cores whereas the MZMs vanishes in the
trivial phase as the magnetic field goes through the TI without
tilting. By changing the tilted angle of the magnetic fields, one
can manipulate the ratio of $t_d/t_s$ to trigger a topological
transition between trivial and topological phases.

{\large \bf Computational methods.} With a finite interaction in
Eq.~(\ref{fermionic basis}), exact characterization of the ground
state is beyond the single-particle picture. Although one can
still perform the Hartree-Fock approximation to decouple the
interaction term as
\begin{align}
n_j n_{j+1} \sim \langle n_j \rangle n_{j+1} + n_j \langle n_{j+1}
\rangle - \langle n_j \rangle \langle n_{j+1}\rangle - \big(
\chi_j c^{\dag}_{j+1}c_j +   c^{\dag}_j c_{j+1} \chi^*_j - |\chi_j
|^2 \big), \nonumber
\end{align}


{\large \bf Physical quantities.} The first signature we use to
identify the MZMs is a zero energy gap between the lowest
even-parity and odd-parity states, $\Delta E \equiv
|E_0(P=1)-E_0(P=-1)|=0$. It reflects Majorana modes occupying two
zero-energy levels, also causing double degeneracy in the ground
state. The IDW phase has non-zero but small $\Delta
E$~\cite{Thomale2013}, while the other trivial phases have
relatively large $\Delta E$. Another signature to characterize the
topological property is to compute the entanglement spectrum. The
entanglement spectra $\lbrace \varepsilon \rbrace$ are simply the
eigenvalues of reduced density matrices
\begin{align}
\rho _l= \textrm{Tr}_r |\psi_0 \rangle \langle \psi_0 |, \nonumber
\end{align}
where the subscript $l$ represents partially tracing out the
degrees of freedom of the right block. The topological phase has
two-fold degeneracies of the entire entanglement spectrum. Rather
than observing the entanglement spectrum, throughout the main
context, we compute the unitless $\delta \varepsilon$ defined as
\begin{align}
\delta \varepsilon = \sum_{P= \pm1} \sum_n
(\varepsilon^P_n -\varepsilon^P_{n+1})^2,
\end{align}
to distinguish the topological from trivial
phases~\cite{Pollmann2010}. The first summation is over the ground
states in two parity sectors. In the topological phase, both the
ground state and the entanglement spectra are doubly degenerate,
so all the paired entanglement spectrum difference
$(\varepsilon^P_n -\varepsilon^P_{n+1})$ vanish and $\delta
\varepsilon=0$. This property is robust even in the presence of
interaction~\cite{Chan2015} and easily implemented with numerical
simulation.

In the large-$U$ limit, the system is a commensurate
charge-density-wave (CDW) insulator, which is topologically
trivial since neither the entanglement spectrum nor the
ground-state energy shows double degeneracy. A CDW state has
electrons residing on every other lattice sites (in a classical
picture) to lower the interaction energy $4Un_j n_{j+1}$, and can
hence be characterized by the structure factor of charge-charge
correlations
\begin{align}
S(q) = \frac{1}{L}\sum_{j',j} \langle n_{j'} n_j \rangle e^{i q
(x_{j'}-x_j)}. 
\end{align}
The CDW order parameter can be defined as $O_{\rm{CDW}} = \lim_{L
\to \infty} \sqrt{S(q)/L}$. In the thermodynamic limit, a finite
$O_{\rm{CDW}}$ implies the existence of the lone-ranged CDW
ordering. At $q=\pi$, the CDW ordering is commensurate, labeled as
CDWI in the phase diagrams Fig.~\ref{fig:phasediagram}. For other
values of $q$, it is incommensurate; in the main context, we have
$q=2k_F$ instability in the incommensurate charge-density-wave
liquid state, where $k_F$ is the Fermi wave vector. In the
spinless chain, half-filling $\bar{n}=0.5$ corresponds to
$k_F=\pi/2$. For $0.5 < \bar{n} < 1$, $k_F=(1-\bar{n})\pi$.

Figures \ref{fig:chargestructure} (a) and (b) show the
(unnormalized) charge structure factor $S(q)$ for (a) $t_{1,2}=0$,
$t_d=1.2t_s$ (b) $t_{1,2}=0.1 t_s$, $t_d=0.4t_s$. In both figures,
at relative weak interaction strength, labeled by the black
squares, no peaks are observed. They are located in the MZM and
TvSC states in (a) and (b), respectively. At moderate interaction,
$S(q)$ develops peaks located roughly at $q=2k_F$. Upon increasing
$U$, the filling $\bar{n}$ decreases and approaches to $0.5$, and
$k_F$ moves toward to $\pi/2$. This feature reveals the $2k_F$
charge instability and characterizes the IDW state. At $q=\pi$,
for (a) $U=1.7t_s$ and for (b) $U=t_s$, the ground state is a
CDWI, and consistently, $\bar{n}\simeq 0.5$ at this moment.

{\large \bf Estimation of Majorana coupling and interaction.} To
estimate the strengths of the physical parameters, we consider the
thin film of topological insulator Bi$_2$Se$_3$ on the top of the
superconductor NbSe$_2$, which is an experimental
realization~\cite{Xu_TI_SC} of the Fu-Kane model. First, the
strengths of $t_s$ and $t_d$ stemming from the coupling of the top
and bottom TI surface states are given by
\begin{align}
t_s,\ t_d \sim G_{\rm{bulk}} e^{-h_{\rm{TI}}/\lambda_{\rm{TI}}}
\sim 2\ \rm{meV},
\end{align}
where the bulk gap $G_{\rm{bulk}}$ of Bi$_2$Se$_3$ is about
$0.3$~eV. The thickness of TI on the superconductor in the recent
experiment is $5$ quintuple layers~\cite{Majorana_Fu_Kane_Jia},
which is about $h_{\rm{TI}}\sim 5$ nm. The decay length in the
vertical direction is given by  the Fermi velocity
($\nu_F=2.2$eV$\cdot \AA$) divided by the bulk gap $\hbar \nu_F
/G_{\rm{bulk}} =\lambda_{\rm{TI}}\sim 1$nm \cite{TI_details}. This
hybridization leads to non-zero energy Majorana fermions residing
on the vortices.

We can estimate the values of $t_1$ and $t_2$ based on the
parameters of the superconductivity, since in the absence of the
superconductivity the Majoranas are delocalized on the top and
bottom layers,  Regardless of superconductor proximity effect
\cite{Chiu_SC_induced_gap}, we use the NbSe$_2$ superconducting
gap $G_{\rm{SC}}\sim 1$meV. The estimated values of the intra-leg
tunnelings are given by
\begin{align}t_1, t_2 \sim
G_{\rm{SC}}e^{-d_{\rm{v}}/\lambda_M}\sim 0.3 \rm{meV},
\end{align}
where the distance between two Majoranas $d_{\rm{v}}$ on the same
surface is about $50$ nm and the decay length of Majorana
hybridization strength on the topological insulator $\lambda_{M}$
is close to the London penetration depth of the superconductor
($40$ nm) when the depth is smaller than $\hbar
\nu_F/G_{\rm{SC}}$~\cite{Chiu2011}. By comparing with $t_d$ and
$t_s$, $t_1$ and $t_2$ can be neglected.

The interaction $U$ for two Majoranas on the top and two Majoranas
on the bottom comes from the Coulomb interaction of two electrons
(holes), each of which is the overlap between two Majorana
wavefunctions; hence, the strength of the interaction $U$ of four
Majorana $\alpha_{1,2,3,4}$ has been written in the density
function $\rho$ of electron
(hole)~\cite{Chiu_Majorana_interaction}
\begin{align}
U=&-8(g_{1234}+g_{4123}-g_{1234}),
\end{align}
where
\begin{align}
g_{ijkl}=&\frac{1}{2}\int \int dr^2 dr'^2
\rho_{ij}(r)V(r-r')\rho_{kl}(r'), \nonumber
\end{align}
where $V(r-r')$ indicates the effective Coulomb potential. Since
the overlap ($e^{-h_{\rm{TI}}/\lambda_{\rm{TI}}}$) between the top
and bottom TI surface Majoranas is less than on the same surface
($e^{-d_{\rm{\rm{v}}}/\lambda_{M}}$), the overlap of Majoranas on
the surface is considered as major contribution to the
interaction, or $\rho \sim e^{-d_{\rm{\rm{v}}}/\lambda_{M}}$. The
reason is that $h_{\rm{TI}}/\lambda_{\rm{TI}}> d_{\rm{v}}/
\lambda_M$, The Coulomb potential, which can be estimated by the
ionization energy of hydrogen $E_H$, is given by
$V=\frac{E_{\rm{H}}}{\epsilon}\frac{a_{H}}{h_{\rm{TI}}}$, where
$a_H$ is the Bohr radius and the dielectric constant $\epsilon$ is
about $20$ due the screening of the Coulomb interaction. The value
of the effective interaction energy is roughly
\begin{align}
U \sim \rho^2 V =
\frac{E_{\rm{H}}}{\epsilon}\frac{a_{H}}{h_{\rm{TI}}}e^{-2d_{\rm{\rm{v}}}/\lambda_{M}}\sim
0.56 ~\rm{meV}. \end{align}
This estimation is in agreement with
Ref.~\cite{Chiu_Majorana_interaction}, which adopted another
method for the estimation. Therefore, comparing the strengths of
the interaction and hopping, we obtain the ratio $U/t_s \sim
0.28$.

\section*{Acknowledgements}

HHH is grateful for helpful discussion with Allan MacDonald. KS is
grateful to Carlos J. Bolech, Nayana Shah, and Chuanwei Zhang for
informative discussions. KS is supported by ARO
(W911NF-12-1-0334), AFOSR (FA9550-13-1-0045), and NSF
(PHY-1505496). JW would like to thank Su Yi, Kai Chang and Zhenyu
Zhang for their helpful discussion. He also wants to thank SUSTC
for the startup funding and Shenzhen Peacock Plan and Shenzhen
Fundamental Research Funds JCYJ20150630145302225. CKC would like
to acknowledge the support of the Max-Planck-UBC Centre for
Quantum Materials, Microsoft and LPS-MPO-CMTC and thank M. Franz
and G. Bian for their helpful discussion. We acknowledge the
computational resource provided by University of Cincinnati and
Texas Advanced Computing Center (TACC).

\section*{Author contributions statement}

H.~H.~Hung performed the numerical calculations and provide the
data for Figures 2--4. C.~K.~Chiu proposed the Fu-Kane model setup
Hamiltonian, performed the analytic calculation, and estimate the
interaction strength. K.~Sun visualized the system by providing
Figure 1 and gave scientific advice in every phase of this work.
J.~Wu transformed the Hamiltonian from Majorana basis to Fermion
basis and drew Figures 2--4. All authors participated in writing
and reviewing the manuscript.

\end{document}